\documentclass[aip,rsi,superscriptaddress,showpacs,amsmath,amssymb]{revtex4-1}

\usepackage{graphicx}
\usepackage[load=prefixed,separate-uncertainty=true,multi-part-units=single,range-phrase=--,range-units=single,list-units=single]{siunitx}
\usepackage[capitalize]{cleveref}
\usepackage{caption}
\usepackage{keyval}
\usepackage{subfig}
\usepackage{microtype}
\usepackage{xcolor}
\usepackage{tensor}
\usepackage{etoolbox}

\robustify{\subref}

\DeclareSIUnit\mJ{\milli\J}

\providecommand{\ani}[1]{\ensuremath{#1^{-}}}
\providecommand{\dani}[1]{\ensuremath{#1^{2-}}}
\providecommand{\cati}[1]{\ensuremath{#1^{+}}}
\providecommand{\iso}[2]{\smash{\ensuremath{^{#1}#2}}}
\providecommand{\mol}[2][]{\ensuremath{\text{#2}_{#1}}}

\begin{document}
	
	\title{Dianion Diagnostics in DESIREE: High-Sensitivity Detection of \dani{\mol[n]{C}} from a Sputter Ion Source}
	\author{K. C. Chartkunchand}
		\email{kiattichart.chartkunchand@fysik.su.se}
		\affiliation{Department of Physics, Stockholm University, AlbaNova, SE-106 91 Stockholm, Sweden}
	\author{M. H. Stockett}
	\author{E. K. Anderson}
	\author{G. Eklund}
	\author{M. K. Kristiansson}
		\affiliation{Department of Physics, Stockholm University, AlbaNova, SE-106 91 Stockholm, Sweden}
	\author{M. Kami\'{n}ska}
		\affiliation{Department of Physics, Stockholm University, AlbaNova, SE-106 91 Stockholm, Sweden}
		\affiliation{Institute of Physics, Jan Kochanowski University, 25-369 Kielce, Poland}
	\author{N. de Ruette}
	\author{M. Blom}
	\author{M. Bj\"{o}rkhage}
	\author{A. K\"{a}llberg}
	\author{P. L\"{o}fgren}
	\author{P. Reinhed}
	\author{S. Ros\'{e}n}
	\author{A. Simonsson}
	\author{H. Zettergren }
	\author{H. T. Schmidt}
		\email{schmidt@fysik.su.se}
	\author{H. Cederquist}
		\affiliation{Department of Physics, Stockholm University, AlbaNova, SE-106 91 Stockholm, Sweden}
	
	\begin{abstract}
	A sputter ion source with a solid graphite target has been used to produce dianions with a focus on carbon cluster dianions, \dani{\mol[n]{C}}, with $n=$ \numrange{7}{24}.  Singly and doubly charged anions from the source were accelerated together to kinetic energies of \SI{10}{\keV} per atomic unit of charge and injected into one of the cryogenic (\SI{13}{\K}) ion-beam storage rings of the DESIREE (Double ElectroStatic Ion Ring Experiment) facility at Stockholm University.  Spontaneous decay of internally hot \dani{\mol[n]{C}} dianions injected into the ring yielded \ani{\mol[n]{C}} anions with kinetic energies of \SI{20}{\keV}, which were counted with a microchannel plate detector.  Mass spectra produced by scanning the magnetic field of a \ang{90} analyzing magnet on the ion injection line reflects the production of internally hot \dani{\mol[7]{C}}--\dani{\mol[24]{C}} dianions with lifetimes in the range of tens of microseconds to milliseconds.  In spite of the high sensitivity of this method, no conclusive evidence of \dani{\mol[6]{C}} was found while there was a clear \dani{\mol[7]{C}} signal with the expected isotopic distribution.  This is consistent with earlier experimental studies and with theoretical predictions.  An upper limit is deduced for a \dani{\mol[6]{C}} signal that is two orders-of-magnitude smaller than that for \dani{\mol[7]{C}}.  In addition, \dani{\mol[n]{C}\mol{O}} and \dani{\mol[n]{C}\mol{Cu}} dianions were detected.
	\end{abstract}
	
	
	\date{\today}
	
	\maketitle
	
	\section{Introduction}
	In this paper, we present a highly sensitive method to probe the production of multiply charged anions utilizing the cryogenic electrostatic ion storage ring DESIREE.  A number of outstanding issues relating to the inherent stabilities and reactivities of such ions exists, which is of clear interest for astrophysical applications.  Examples include the \dani{\mol[n]{C}} carbon cluster dianions, and in particular the thermodynamic stability~\cite{Sommerfeld93} of \dani{\mol[7]{C}}, as well as the stabilities of \dani{\mol[60]{C}} and other fullerene dianions~\cite{Liu04,Hampe02}.
		
	Small and large carbon-bearing molecules are likely to be important for various processes in space.  There, they are often ionized and may also carry extra electrons, i.e., they may be positively or negatively charged.  Examples of the latter are \ani{\mol[4]{C}\mol{H}}, \ani{\mol[6]{C}\mol{H}}, \ani{\mol[8]{C}\mol{H}}, \ani{\mol{CN}}, \ani{\mol[3]{C}\mol{N}}, and \ani{\mol[5]{C}\mol{N}}, which were discovered through detailed comparisons between laboratory rotational spectra and astronomical observations~\cite{Cernicharo07,McCarthy06,Brunken07,Thaddeus08,Cernicharo08,Agundez10}.  Similarly sized pure carbon cluster anions, \ani{\mol[n]{C}}, are also believed to play important roles in space but have so far not been detected.  Astrophysical carbon cluster anions are likely to be formed ``hot,'' i.e., with high internal excitation energies, through the attachment of free electrons~\cite{Herbst81,Larsson12}.  Recently, large differences between cooling rates of internally hot molecular ions have been reported and discussed in terms of so-called recurrent fluorescence processes~\cite{Ito14,Chandrasekaran14,Kono15,Ebara16,Martin13}, which should be an important consideration when modeling astrochemical reaction networks~\cite{Walsh09}.  Other carbon-bearing molecules, such as neutral \mol[60]{C} and \mol[70]{C} fullerenes, have been observed in planetary nebulae~\cite{Cami10,GarciaHernandez10,Sellgren10,Berne11}, and a number of \cati{\mol[60]{C}} absorption bands have been identified with some of the so-called diffuse interstellar bands (DIBs)~\cite{Campbell15,Walker15,Campbell16,Walker17,Spieler17}.  In addition, polycyclic aromatic hydrocarbons (PAHs) or other fullerenes, as well as their ions, have been suggested as carriers of other DIB features~\cite{Cami10,Tielens13}.  For astrophysical applications, there is thus a clear need to find efficient ways to produce and study carbon-based molecular anions. 
	
	\emph{Multiply charged} anions may be present in dark interstellar clouds and could also play a role in chemical reaction networks.  However, up until now the inherent stabilities of such systems have been very difficult to study.  This is due to difficulties in producing sufficient quantities of such ions and to low electron emission rates, as even thermodynamically unstable systems may have high-energy barriers against electron emission.  Many measurements are also complicated by the influence of the blackbody radiation field, which may be substantial in typical room-temperature experiments.  This situation is now rapidly improving as techniques to cryogenically store and analyze singly and multiply charged anions and cations in electrostatic ion-beam storage rings and traps have become available~\cite{Wolf15}.

	Cryogenic electrostatic ion-beam storage devices allow for the storage of ions of any mass at high velocities under superb vacuum conditions, with only \numrange{e2}{e4} rest gas molecules~\cite{Schmidt13,vonHahn16} per \si{\cmc}. This opens the possibility for action spectroscopy studies (see for examples Refs.~\cite{Nielsen01,Kiefer16,Spieler17} for descriptions of this method) and the counting of individual reaction products~\cite{Thomas11,vonHahn11}.  So far, only three cryogenic electrostatic ion-beam storage rings are in operation worldwide. These are the 35-meter circumference Cryogenic Storage Ring (CSR) at the Max Planck Institute in Heidelberg, Germany~\cite{vonHahn11,vonHahn16}, the three-meter circumference RIKEN Cryogenic Electrostatic (RICE) ring at the RIKEN laboratory in Tokyo, Japan~\cite{Nakano12,Nakano17}, and the DESIREE rings at Stockholm University in Sweden~\cite{Thomas11,Schmidt13}. The latter has a unique double-ring configuration, with two 8.6-meter circumference rings sharing a common straight section for merged anion-cation reaction studies~\cite{OConnor16,Meyer17,Schmidt17}. A few cryogenic electrostatic ion traps are also in operation~\cite{Schmidt01,Lange10,Rubinstein17} where similar studies may also be performed. 

	In the present study, one of the DESIREE storage rings is used to diagnose the production of beams of carbon cluster dianions from a sputter ion source.  While very short-lived ($\sim$\SI{1}{\fs}) \dani{\mol[2]{C}} dianions have been observed~\cite{Andersen96}, \dani{\mol[7]{C}} is the smallest carbon cluster dianion for which thermodynamical stability is predicted~\cite{Sommerfeld93}.  This has so far not been investigated in detail experimentally, although metastable and/or stable \dani{\mol[7]{C}} ions have previously been observed~\cite{Schauer90,Gnaser93,Calabrese96,Middleton97,Franzreb05}.  The lack of evidence for \dani{\mol[5]{C}} and \dani{\mol[6]{C}} in previous experiments~\cite{Franzreb05,Schauer90,Middleton97} also seems to support \dani{\mol[7]{C}} as the smallest thermodynamically stable carbon cluster dianion.  A related open issue concerns the thermodynamic stability of small fullerene dianions such as \dani{\mol[60]{C}} and \dani{\mol[70]{C}}, which so far have only been investigated using room-temperature storage devices~\cite{Liu04,Tomita06,Kadhane09,Hampe02,Lassesson05} and a radio-frequency trap~\cite{Wang06} cooled to \SI{70}{\K}.  \emph{Cryogenic} storage rings have a clear advantage over the room-temperature devices as they can be used to follow decay processes on much longer timescales~\cite{Backstrom15}.
	
	In Section II, the detection technique developed for DESIREE, which can be applied to any dianion production method, will be described.  In Section III, we will show the resulting pure dianion mass spectra and compare them with the much higher-intensity mass spectra of anions also produced in the source.  A weak signal at the position expected for \dani{\mol[6]{C}} in the dianion spectrum is present, but we show that this signal is primarily due to secondary particles from the \ani{\mol[3]{C}} beam.

	\section{Experimental Apparatus}
	\begin{figure*}[htp]
		\centering
		\includegraphics[width=\linewidth]{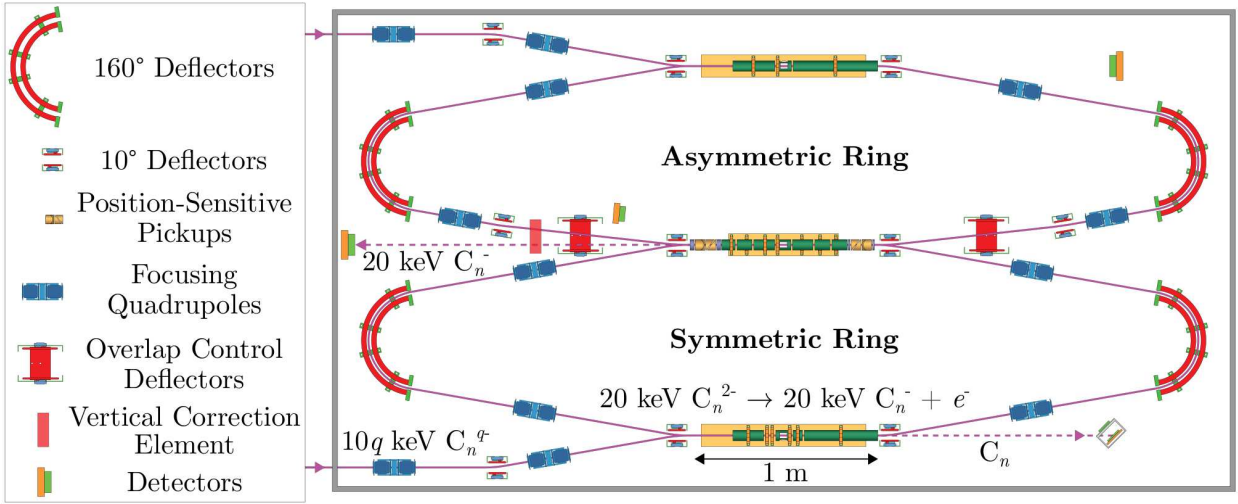}
		\caption{Schematic of the DESIREE storage rings.  Carbon cluster anions are injected into the Symmetric Ring, which is configured to only pass \SI{20}{\keV} \ani{\mol[n]{C}} anions resulting from spontaneous decay of \SI{20}{\keV} \dani{\mol[n]{C}} dianions within the first straight section of the ring (see text for details).  Only the Vertical Correction Element (VCE) used in this experiment is shown; other such elements are present in the ring but omitted in the figure for clarity.}
		\label{DESIREE}
	\end{figure*}
	Beams of carbon-containing anions were produced using a SNICS cesium-sputter ion source~\cite{SNICS} with a solid graphite target mounted in a copper cathode.  The ions were accelerated to \SI{10}{\keV} per atomic unit of charge and their mass-to-charge ratios selected by means of a \ang{90} double-focusing dipole magnet.  The field in this analyzing magnet was measured by means of a LakeShore Model 460 3-Channel Hall-effect Gaussmeter and this reading was used to calibrate the mass-to-charge ratio scales in the anion and dianion mass spectra as described below. After the magnet, ions with a well-defined mass-to-charge ratio were transported to the injection port of the so-called Symmetric Ring of DESIREE, shown in~\cref{DESIREE}.  At this point the ions have traveled a distance of \SI{10.6}{\m} from the ion source to the injection port.  Singly and doubly charged carbon cluster anions in a beam with a certain mass-to-charge ratio, $m/q$, have the same velocity but different kinetic energies of 10 and 20 \si{\keV}, respectively. As an example, a \SI{20}{\keV} \dani{\mol[7]{C}} ion travels \SI{1}{\m} in about \SI{5}{\mics}. Taking into account the time for transport at lower energy between the source and the \SI{10}{\keV} acceleration stage, the \dani{\mol[n]{C}} ions typically reach the injection stage of the ring in hundreds of microseconds after leaving the sputter source. This means that dianions will only reach the ring if they decay on comparable or longer timescales, or if they are stable. The deflection plates in the injection stage were set for injection of ions with $10q$ \si{\keV} kinetic energy, $q$ being the charge state of the ion.  Thus, \SI{10}{\keV} anions $(q=1)$ and \SI{20}{\keV} dianions $(q=2)$ with the same mass-to-charge ratios enter the ring as parts of the same ion beam.  For measurements of dianion mass spectra, the voltages on the deflectors of the first \ang{180}-bend, consisting of two \ang{10}-bends and one cylindrical \ang{160}-bend as shown in \cref{DESIREE}, were set to only admit ions with a kinetic energy of \SI{20}{\keV} per atomic unit of charge. That is, only those \dani{\mol[n]{C}} ions that lose one of their extra electrons in the first straight section of the Symmetric Ring between the injection stage and the entrance of the first \ang{180}-bend can continue as \SI{20}{\keV} \ani{\mol[n]{C}} ions through this bend.  After this they continue straight to the detector positioned after the second straight section of the ring (see~\cref{DESIREE}) and are counted by means of a position-sensitive microchannel plate (MCP) detector operated at cryogenic temperatures~\cite{Rosen07}.  The voltages of the second \ang{180}-bend in the Symmetric Ring were set to zero.  Mass spectra were also recorded by measuring the ion current in a Faraday cup (FC) after the exit slit of the analyzing magnet.  This current was dominated by \SI{10}{\keV} \ani{\mol[n]{C}} ions, but also contained small amounts of \SI{20}{\keV} \dani{\mol[n]{C}} ions.  Separate mass spectra of dianions and anions were measured by recording the count rate of dianions on the MCP detector and by recording the FC current as a function of the \ang{90} analyzing magnet field settings, with an overall mass resolution of $m/\Delta m\sim350$.
	\begin{figure*}[htp]
		\centering
		\includegraphics[width=\linewidth]{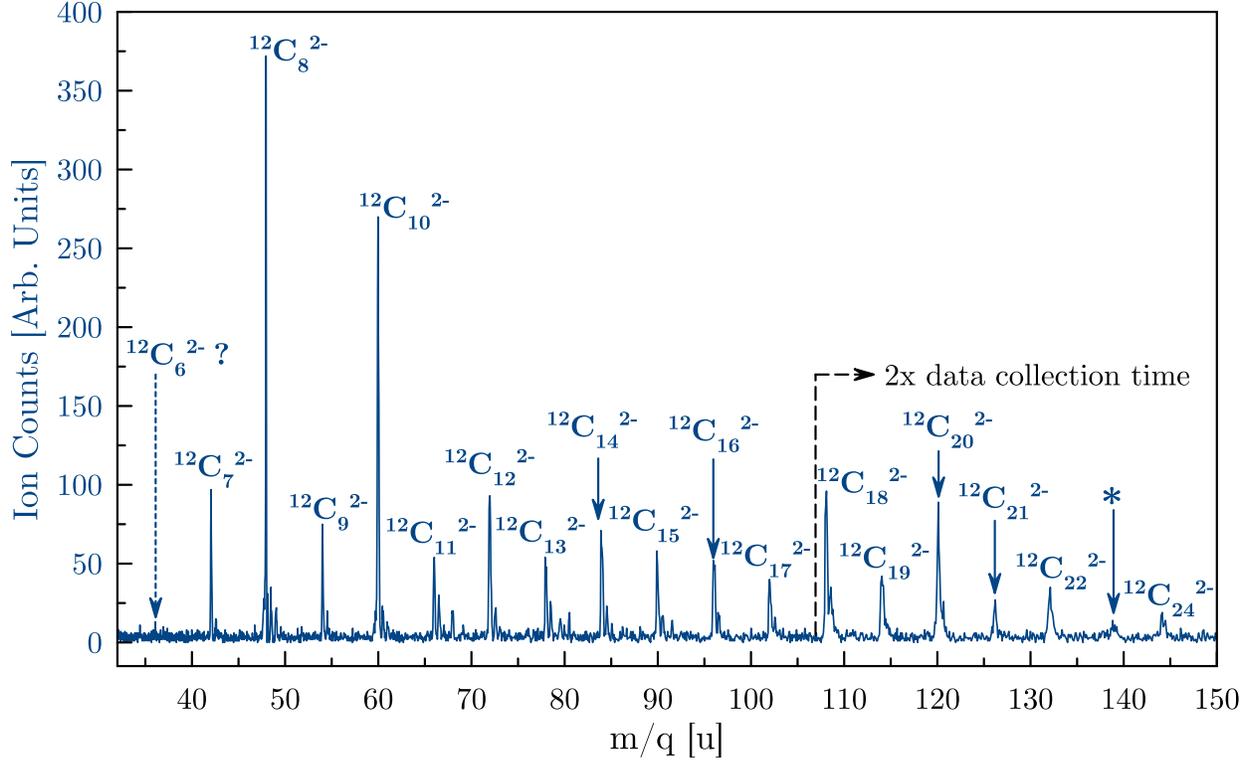}
		\caption{The \dani{\mol[n]{C}} mass spectrum recorded according to the procedures in the text.  Twice the counting time was used for the $m/q=$ \SIrange{106}{150}{\amu} segment of the mass scan.  The dashed arrow labeled ``\dani{\iso{12}{\mol[6]{C}}} ?'' indicates the expected location of this ion in the dianion spectrum.  Note that the observed signal is from those dianions that have lost one of their additional electrons, and that the relative intensities of the observed peaks are \emph{not} simply proportional to the probability of dianion formation.  The peak labeled with an asterisk indicates the presence of an unknown dianionic species close to the expected location of the \dani{\iso{12}{\mol[23]{C}}} peak (see discussions in the main text).}
		\label{C--Scan}	
	\end{figure*}	
	\section{Results and Discussion}
\begin{figure}
		\includegraphics[width=\linewidth]{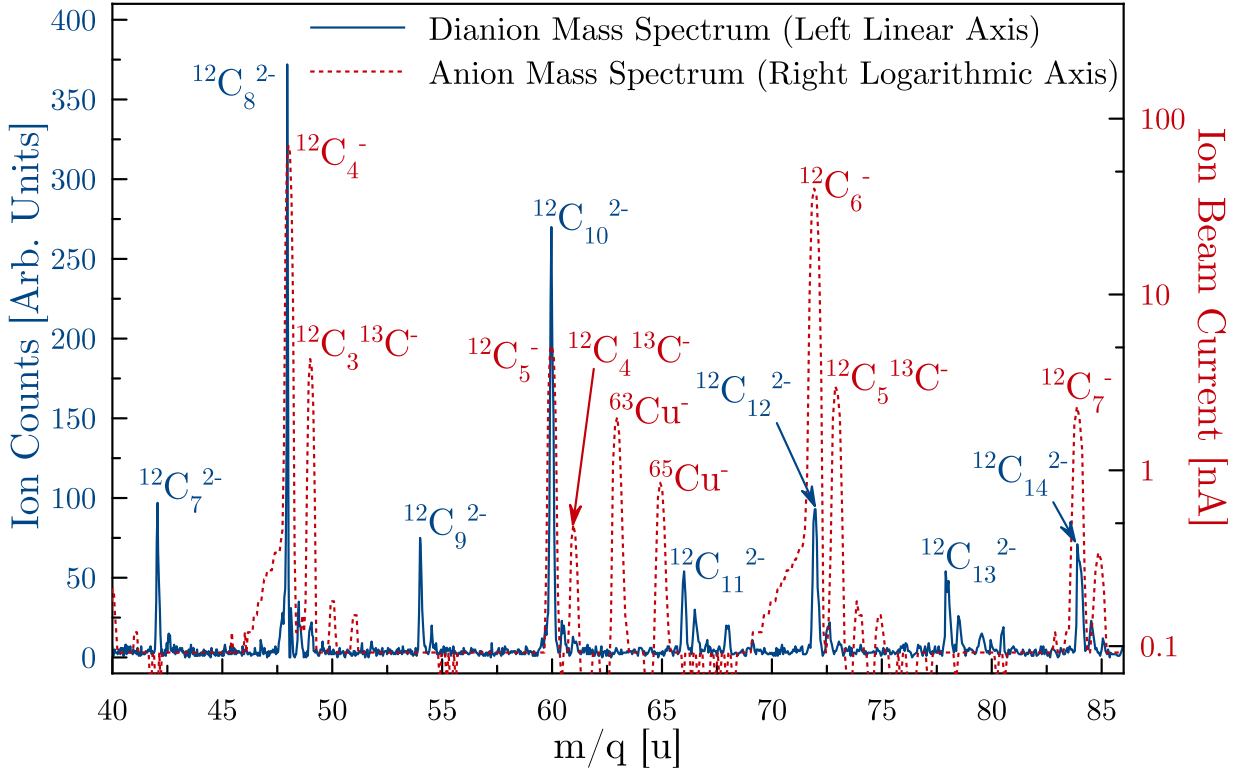}
		\caption{The \dani{\mol[n]{C}} mass spectrum (in blue) overlaid with the \ani{\mol[n]{C}} mass spectrum in the range $m/q=$ \SIrange{40}{87}{\amu}.  Note the logarithmic scale for the ion beam current corresponding to the \ani{\mol[n]{C}} mass scan (in red).}
		\label{Diani+Mass}
	\end{figure}

	Mass spectra for \dani{\mol[n]{C}} and \ani{\mol[n]{C}} are shown in~\cref{C--Scan,Diani+Mass}, respectively.  \cref{C--Scan} shows \dani{\mol[n]{C}} dianions in the range $m/q=$ \SIrange{32}{150}{\amu}, corresponding to the expected positions for dianions with $n=$ \numrange{6}{24}, while~\cref{Diani+Mass} shows \dani{\mol[n]{C}} dianion peaks overlaid with \ani{\mol[n]{C}} anion peaks in the range $m/q=$ \SIrange{40}{85}{\amu}.  The pure \dani{\mol[n]{C}} mass spectrum clearly exhibits an odd-even alternation in intensities, which has been observed in previous measurements~\cite{Schauer90,Gnaser93,Middleton97}.  Mass peaks are clearly defined for carbon cluster dianions up to $n=17$.  Peaks above this are less well-defined, making it more difficult to determine whether they are due to pure \dani{\mol[n]{C}} clusters or some mixture of other dianionic species.  In particular, the peak in between the \dani{\iso{12}{\mol[22]{C}}} and \dani{\iso{12}{\mol[24]{C}}} peaks shown in~\cref{C--Scan} occurs at $m/q=$ \SI{138.5}{\amu} instead of \SI{138}{\amu} as would be expected for \dani{\iso{12}{\mol[23]{C}}}.  This can be explained by the presence of an unknown, doubly-charged impurity that dominates over any \dani{\iso{12}{\mol[23]{C}}} that may have been injected into the ring.  In~\cref{Diani+Mass}, the dianion mass spectrum was recorded from particle counts on the MCP detector (left vertical scale, blue spectrum), while the anion spectrum was recorded from FC current readings (right vertical scale, red spectrum); note the logarithmic vertical scale of the anion spectrum (in red).  While the anion spectrum reflects the relation between the different anion production efficiencies in the sputter source under the present operating conditions, the interpretation of the dianion mass spectrum is less straightforward. The intensities of the dianion peaks depend on the efficiencies with which dianions are produced in the source but also on the $\dani{\mol[n]{C}}\rightarrow\ani{\mol[n]{C}}+\ani{e}$ decay rate in a rather involved way. If this rate is high, the dianions decay before reaching the ring; if it is low, it is unlikely that spontaneous decay occurs in the first straight section of the Symmetric Ring and the signal becomes very weak.  For sufficiently cold, thermodynamically stable \dani{\mol[n]{C}} dianions, there is no signal apart from possible very minor contributions due to electron stripping in the extremely dilute residual gas density of the ring (\numrange{e3}{e4} \mol[2]{H} molecules per \si{\cmc}).  Such dianions could be detected through photodetachment with a laser introduced co-linearly to the ion beam  along the straight section on the ion-injection-side of the ring.  In this case, \ani{\mol[n]{C}} produced through single-electron detachment would pass through the first \ang{180}-bend and be detected, indicating the presence of \dani{\mol[n]{C}} injected into the ring.  However, this has not been implemented for the present pilot study.

	Anions and dianions are produced hot and with broad internal energy distributions in the sputter ion source. Typically vibrational and rotational temperatures are several thousand Kelvin, as shown in previous studies of anion relaxation processes in ion-beam storage rings~\cite{Hansen17,Menk14} and traps~\cite{Breitenfeldt16}.  In many cases, these properties are manifested as non-exponential decays of neutralization signals due to spontaneous decay of the ions as functions of the storage time $t$.  This type of decay instead often follows a $t^{-1+\delta}$ trend, i.e., a power law behavior in which $\delta$ is often, but not always, small~\cite{Hansen01}.  Depending on the vibronic structure of the anion, recurrent florescence can also become a significant relaxation process with a near-exponential decay of the signal strength over time~\cite{Ito14,Chandrasekaran14,Kono15,Ebara16,Martin13}.  In the present experiment, only those dianions that are produced in reasonable amounts in the source and which decay on timescales that are neither too short nor too long can contribute significantly to the signal.

\begin{figure}[htp]
		\centering
		\includegraphics[width=\linewidth]{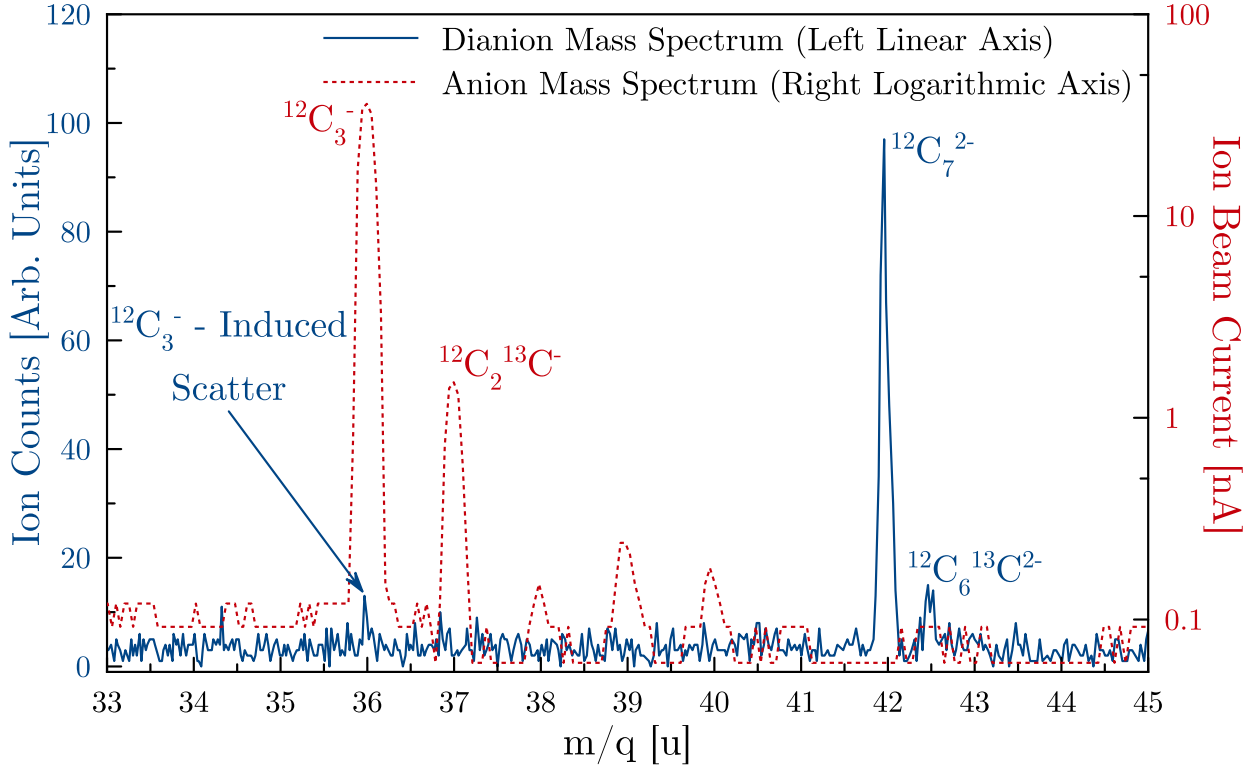}
		\caption{\dani{\mol[n]{C}} and \ani{\mol[n]{C}} mass spectra in the range $m/q=$ \SIrange{33}{45}{\amu}, including \dani{\mol[7]{C}} and any potential signal due to \dani{\mol[6]{C}}.}
		\label{C6C7Zoom}
	\end{figure}

\begin{figure}[htp]
		\centering
		\includegraphics[width=\linewidth]{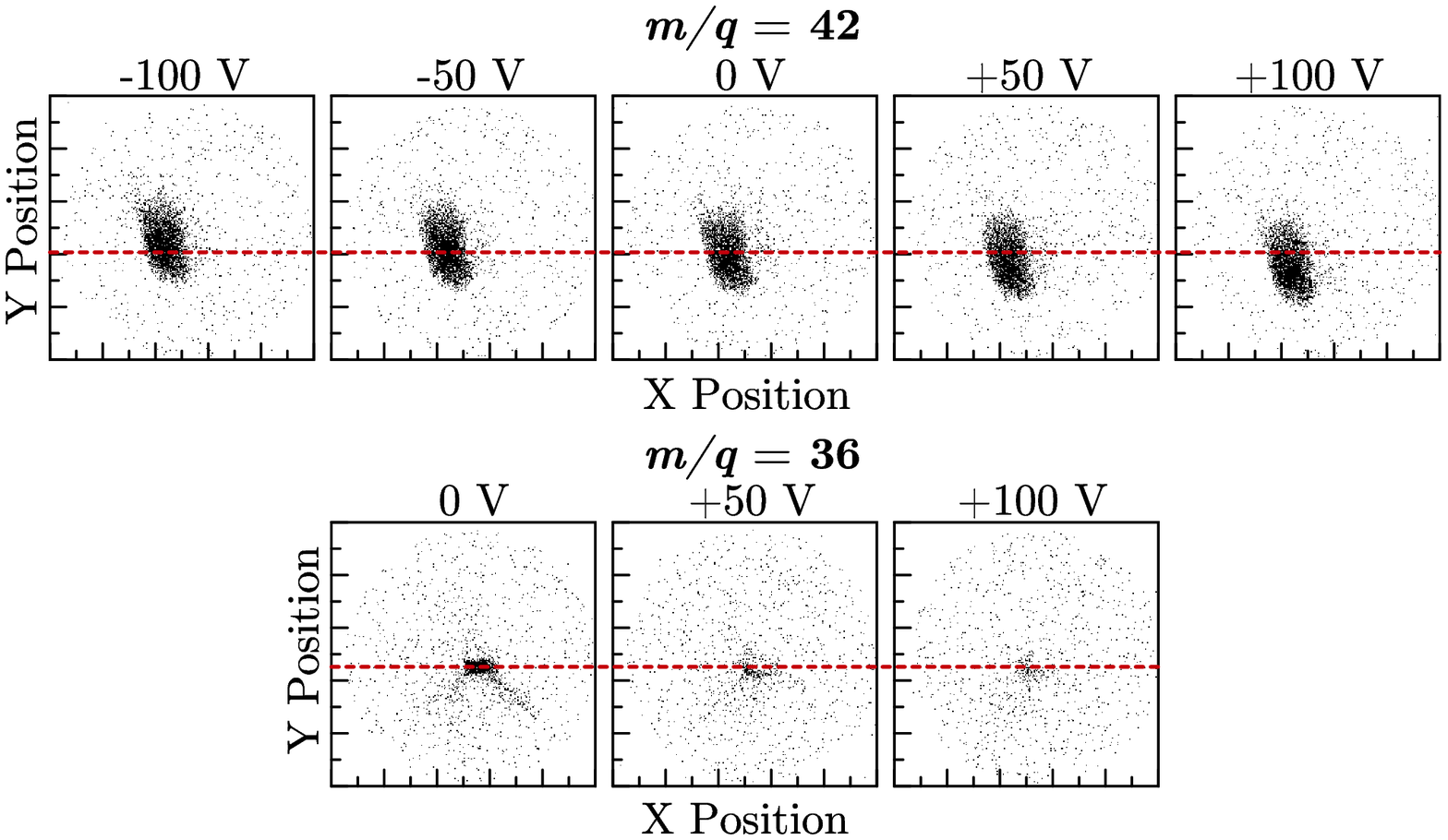}
		\caption{Ion beam positions on the MCP detector for $m/q=$ \SI{42}{\amu} and $m/q=$ \SI{36}{\amu} (corresponding to \dani{\mol[7]{C}} and \dani{\mol[6]{C}}, respectively) for different voltages applied to the Vertical Correction Element (VCE).  The dashed line overlaid on both sets of images corresponds to the vertical center of the beam position with \SI{0}{\V} applied to the VCE.}
		\label{Position}
	\end{figure}

	In~\cref{C6C7Zoom}, a zoom-in is shown of the \dani{\mol[7]{C}} group of peaks and the mass region expected for a possible \dani{\mol[6]{C}} contribution. The intensity distribution of the \dani{\mol[7]{C}} peaks is consistent with the one expected from the natural \iso{13}{\mol{C}} abundance of 1.1\%.  The appearance of peaks at half-integer mass numbers, in this case at \SI{42.5}{\amu}, is a unique feature of dianion mass spectra and allows for the unambiguous identification of \dani{\iso{12}{\mol[6]{C}}\iso{13}{\mol{C}}} and thus also \dani{\iso{12}{\mol[7]{C}}}.  In terms of \dani{\mol[6]{C}}, there is a very weak peak which consistently appeared at the expected \dani{\iso{12}{\mol[6]{C}}} position.  However, this peak seems to be due to other particles produced through the interaction of the rather strong \ani{\mol[3]{C}} beam with electrode surfaces in the Symmetric Ring.  This was concluded after applying small voltages to the Vertical Correction Element (VCE) located before the MCP detector (see~\cref{DESIREE}). The position of the beam spot on the detector for the $m/q=42$ signal (\dani{\mol[7]{C}}) could be moved vertically with these voltages, while the corresponding spot due to the $m/q=36$ signal (\dani{\mol[6]{C}}) was merely reduced in intensity by the same deflection fields (see~\cref{Position}).  While the exact nature of the signal recorded at $m/q=36$ is unknown, it clearly does \emph{not} behave like a proper ion beam traveling through the ring and hence is most likely not due to any \SI{20}{\keV} \ani{\mol[6]{C}} ions that would indicate the presence of \dani{\mol[6]{C}} injected into the ring.  This, however, does \emph{not} prove that \dani{\mol[6]{C}} does not exist as a stable or metastable ion, only that our sputter ion source operating under the present conditions was unable to produce it in sufficient amounts and in states with suitable lifetimes.
	
\begin{figure}[htp]
		\centering
		\includegraphics[width=\linewidth]{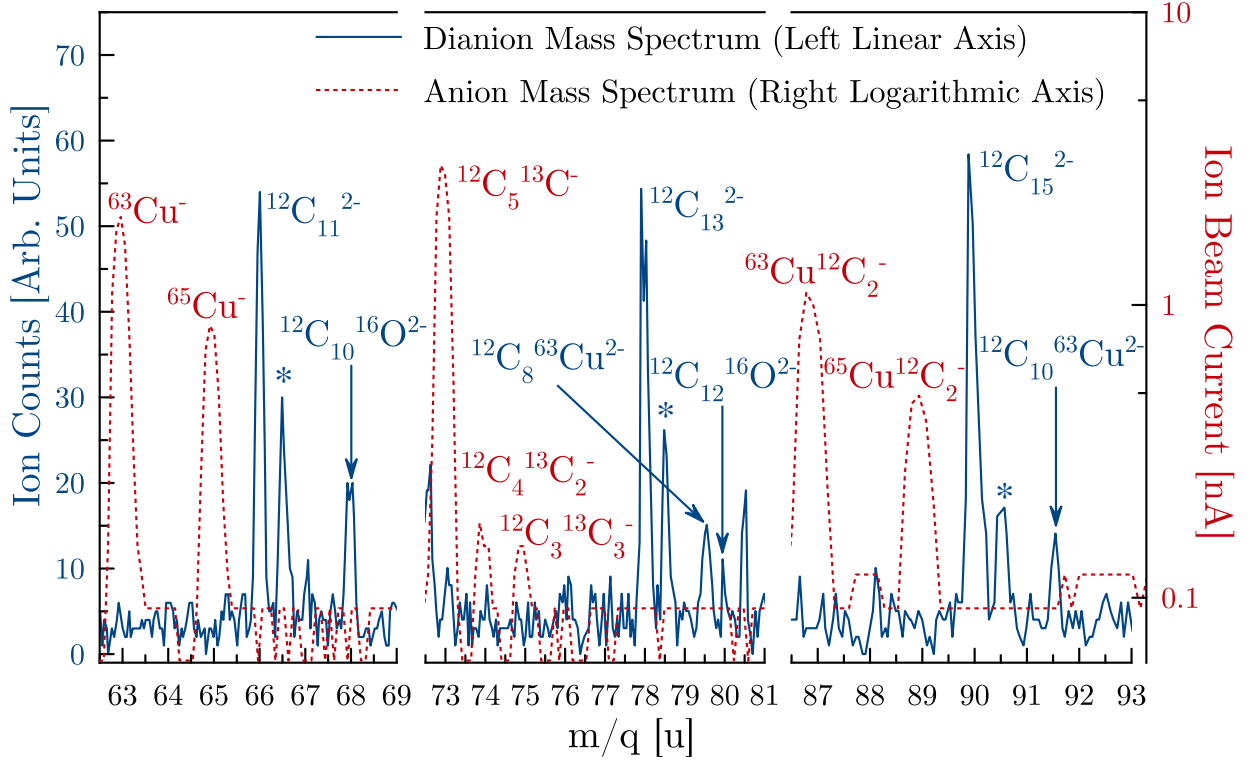}
		\caption{Dianion and anion mass spectra in the \dani{\mol[11]{C}}, \dani{\mol[13]{C}}, and \dani{\mol[15]{C}} regions of interest.  Peaks marked with an asterisk (*) indicate those resulting from substitution of one \iso{13}{\mol{C}} atom into \dani{\mol[11,13,15]{C}}, along with other dianionic species of the same $m/q$ (see discussion in main text).}
		\label{C111315--}
	\end{figure}

	Finally, we remark on the sensitivity of our technique for dianion detection.  In~\cref{C111315--}, both dianion and anion mass spectra in the regions around the \dani{\mol[11]{C}}, \dani{\mol[13]{C}}, and \dani{\mol[15]{C}} peaks are shown.  In these mass spectra, we first note the presence of \dani{\mol[n]{C}\mol{O}} dianions with $n=10,12$.  Dianions of the type \dani{\mol[n]{C}\mol{O}} have been observed previously when oxygen was actively introduced into the sputter ion source~\cite{Gnaser02}.  No such enhancement of \dani{\mol[n]{C}\mol{O}} production was done in the present experiment, demonstrating the sensitivity of our technique.  In regards to the \dani{\mol[11,13,15]{C}} peaks, we also note that the peak intensities for substitution of one \iso{13}{\mol{C}} atom are \emph{greater} than expected based on the natural abundance of \iso{13}{\mol{C}}.  Since there are no indications in the anionic mass spectra of anionic species at the appropriate $m/q$ and as these peaks occur at half-integer mass numbers, we conclude that the additional contribution to these peaks come from other dianionic species.  Possible such dianions could include \dani{\mol[11,13,15]{C}\mol{H}} carbon cluster hydrides~\cite{Franzreb05} or mixed cesium-carbon clusters \dani{\mol{Cs}\mol[2,4]{C}}; the latter is consistent with the ease with which \ani{\mol{Cs}\mol[n]{C}} clusters are produced in cesium sputter sources~\cite{Middleton97,Gnaser99}.  A mixed cesium-carbon cluster dianion, in particular \dani{\mol{Cs}\mol[12]{C}}, may in fact be responsible for the unknown dianion peak at $m/q=$ \SI{138.5}{\amu} seen in the mass spectrum shown in~\cref{C--Scan}.  Similar arguments may also apply to several other peaks in the vicinities of the \dani{\mol[11,13,15]{C}} peaks.  For instance, we note that the mass peaks at \SI{79.5}{\amu} and \SI{91.5}{\amu} have the same $m/q$ as \dani{\iso{12}{\mol[8]{C}}\iso{63}{\mol{Cu}}} and \dani{\iso{12}{\mol[10]{C}}\iso{63}{\mol{Cu}}}, respectively.  Since a graphite target and copper cathode are used in the cesium sputter source only carbon-, cesium-, oxygen-, and copper-containing anions are produced, making this a reasonable assignment for these two mass peaks.  These dianions have, to our knowledge, not been detected previously.  

	\section{Conclusions and Outlook}
	We have demonstrated a highly sensitive method to investigate dianion production methods and their efficiencies. By using the Symmetric Ring of the DESIREE facility as a kinetic-energy-per-charge analyzer, carbon cluster dianion and anion mass spectra could be measured separately, opening new ways to detect small dianion contributions in DESIREE.  This method was used to investigate if metastable \dani{\mol[6]{C}} could be produced in a cesium-sputter ion source.  We determined an upper limit for the corresponding signal rate to be two orders-of-magnitude smaller than that for \dani{\mol[7]{C}} with the present ion source conditions and diagnostic method.  Other dianionic species such as \dani{\mol[n]{C}\mol{Cu}} were also detected without any special enhancement for their production.  Plans are under way to store carbon cluster and fullerene dianion beams in DESIREE and to monitor $\dani{\mol[n]{C}}\rightarrow\ani{\mol[n]{C}}+\ani{e}$ decays as functions of time after injection in order to investigate the inherent stabilities of these \dani{\mol[n]{C}} dianions, as well as to measure their decay lifetimes and binding energies.  Alternative and colder dianion production methods will be tested using the diagnostic method presented here.  These include electrospray ionization techniques, charge exchange in a cesium-vapor cell, and/or pre-cooling in radio frequency buffer-gas pre-traps operating at low temperatures. 
	
	\section*{Acknowledgments}
	This work was performed at the Swedish National Infrastructure, DESIREE (Swedish Research Council Contract No. 2017-00621).  Further support was provided by the Swedish Research Council (Contract No. 821-2013-1642, No. 621-2015- 04990, No. 621-2014-4501, No. 621-2016-06625, No. 621-2016-04181, and No. 2016-03675) and the Knut and Alice Wallenberg Foundation through an earlier DESIREE-investment grant.  M. K. acknowledges financial support from the Mobility Plus Program (Project No. 1302/MOB/IV/2015/0) funded by the Polish Ministry of Science and Higher Education.
	
	\bibliography{References}
	
\end{document}